\UseRawInputEncoding
\pdfoutput=1
\documentclass[conference]{IEEEtran}
\usepackage{amsmath,amsfonts}
\usepackage{algorithmic}
\usepackage{algorithm}
\usepackage[caption=false,font=normalsize,labelfont=scriptsize,textfont=scriptsize]{subfig}
\usepackage{textcomp}
\usepackage{stfloats}
\usepackage{url}
\usepackage{verbatim}
\usepackage{graphicx}
\usepackage{booktabs}
\usepackage{cite}
\begin{document}

\title{DRL-Based Orchestration of Multi-User MISO \\ Systems with Stacked Intelligent Metasurfaces}\vspace{-0.5cm}

\author{\IEEEauthorblockN{Hao Liu\IEEEauthorrefmark{1},
Jiancheng An\IEEEauthorrefmark{2}, Derrick Wing Kwan Ng\IEEEauthorrefmark{3},
George C. Alexandropoulos\IEEEauthorrefmark{4}, and Lu Gan\IEEEauthorrefmark{1}}
\IEEEauthorblockA{\IEEEauthorrefmark{1}School of Information and Communication Engineering, University of Electronic Science and Technology of China, \\Chengdu, Sichuan 611731, China\\
\IEEEauthorrefmark{2}School of Electrical and Electronics Engineering, Nanyang Technological University, Singapore 639798, Singapore\\
\IEEEauthorrefmark{3}School of Electrical Engineering and Telecommunications, University of New South Wales, Sydney, NSW 2052, Australia\\
\IEEEauthorrefmark{4}Department of Informatics and Telecommunications, National and Kapodistrian University of Athens, Athens 15784, Greece\\
Email: \{liu.hao, ganlu\}@uestc.edu.cn, jiancheng.an@ntu.edu.sg, w.k.ng@unsw.edu.au, alexandg@di.uoa.gr
}}
\vspace{-0.5cm}

\maketitle
\vspace{-0.5cm}
\begin{abstract}
Stacked intelligent metasurfaces (SIM) represents an advanced signal processing paradigm that enables over-the-air processing of electromagnetic waves at the speed of light. Its multi-layer structure exhibits customizable increased computational capability compared to conventional single-layer reconfigurable intelligent surfaces and metasurface lenses. In this paper, we deploy SIM to improve the performance of multi-user multiple-input single-output (MISO) wireless systems with low complexity transmit radio frequency (RF) chains. In particular, an optimization formulation for the joint design of the SIM phase shifts and the transmit power allocation is presented, which is efficiently solved via a customized deep reinforcement learning (DRL) approach that continuously observes pre-designed states of the SIM-parametrized smart wireless environment. The presented performance evaluation results showcase the proposed method’s capability to effectively learn from
the wireless environment while outperforming conventional precoding schemes under low transmit power conditions. Finally, a whitening process is presented to further augment the robustness of the proposed scheme.
\end{abstract}


\section{Introduction}
\IEEEPARstart{T}{he} evolution of wireless networks strives to achieve higher transmission rates and lower latency. Next generation mobile networks' advent stimulates advanced physical-layer technologies like reconfigurable intelligent surfaces (RIS) \cite{jian2022reconfigurable, an2022lowcomplexity, an2022codebook, jia2023codebook, liu2023kmean_ris_modulation}, enabling up to $10\times$ data rate improvements \cite{samdanis2020road}. However, in typical wireless networks, the issue of multi-user interference constitutes a significant obstacle that considerably degrades system performance \cite{alexandropoulos2016advanced}. This, in turn, jeopardizes the quality-of-service (QoS) and overall system spectral efficiency. Thus, the need for effective interference cancellation has prompted extensive research attention, spurring advancements in diverse interference management techniques \cite{alexandropoulos2016advanced, hu2021robust}.

Conventional interference cancellation techniques rely on advanced digital signal processors. Very recently, inspired by the wave-based computation \cite{yang2022next}, an all-optical diffractive deep neural network ($\text{D}^2 \text{NN}$) was introduced \cite{sci}, which consists of several passive diffractive layers. Specifically, the $\text{D}^2 \text{NN}$ can compute complex functions and process electromagnetic (EM) signals at the speed of light, substantially reducing energy consumption and processing delay compared to conventional digital signal processors. Nevertheless, the $\text{D}^2 \text{NN}$ is made of three-dimensional printed neurons, constrained to solving a single task once fabricated. To address this issue, a programmable $\text{D}^2 \text{NN}$ \cite{liu2022programmable} was reported, using stacked intelligent metasurfaces (SIM), which can adjust the network coefficients in real-time through an intelligent controller. The SIM inherits the advantages of $\text{D}^2 \text{NN}$ for light-speed network operation while retaining retrainability across diverse machine learning tasks.

SIM has the potential to supplant baseband beamforming in point-to-point communication systems \cite{an2023tutoral, an2023stacked_6g, an2023stacked_MIMO_transsciver, nadeem2023hybrid}. Specifically, \emph{An et al.} \cite{an2023stacked} discussed the application of SIM to multi-user beamforming in a multiple-input single-output (MISO) system, and presented an alternating optimization (AO) approach to solve the non-convex design optimization problem. However, conventional AO approaches are unstable, stemming from acute sensitivity to initial values, parameters, and problem structure. In contrast, deep reinforcement learning (DRL) thrives by leveraging neural networks for robust feature extraction and adaptive learning algorithms that dynamically adjust to evolving environments and tasks \cite{alex2022ml_rl}. Motivated by this potential, in this paper, we present a SIM-assisted multi-user downlink MISO communication system optimized by DRL. In the considered system, the SIM is assigned to perform the wave-based precoding with low complexity transmit radio frequency (RF) chains. We then formulate an optimization problem that jointly optimizes the coefficients of the SIM meta-atoms and the power allocation strategy of transmit antennas, which is dynamically solved via a novel DRL approach. The presented numerical results unveil the advantages of the proposed SIM-assisted multi-user MISO wireless communication system across various setups.

\vspace{-0.1cm}
\section{Modeling and Problem Formulation}
\vspace{-0.1cm}
In this section, we introduce the model of the SIM-aided multi-user MISO system in which the SIM acts as the wave-based precoder. Specifically, Section II-A presents the proposed SIM design, while Section II-B introduces the spatially correlated channel model. Finally, Section II-C outlines the considered system performance metric and includes the optimization problem formulation.
\vspace{-0.1cm}
\subsection{SIM Design}
\vspace{-0.1cm}
\begin{figure}[!t]
\centering
\includegraphics[width=0.8\linewidth]{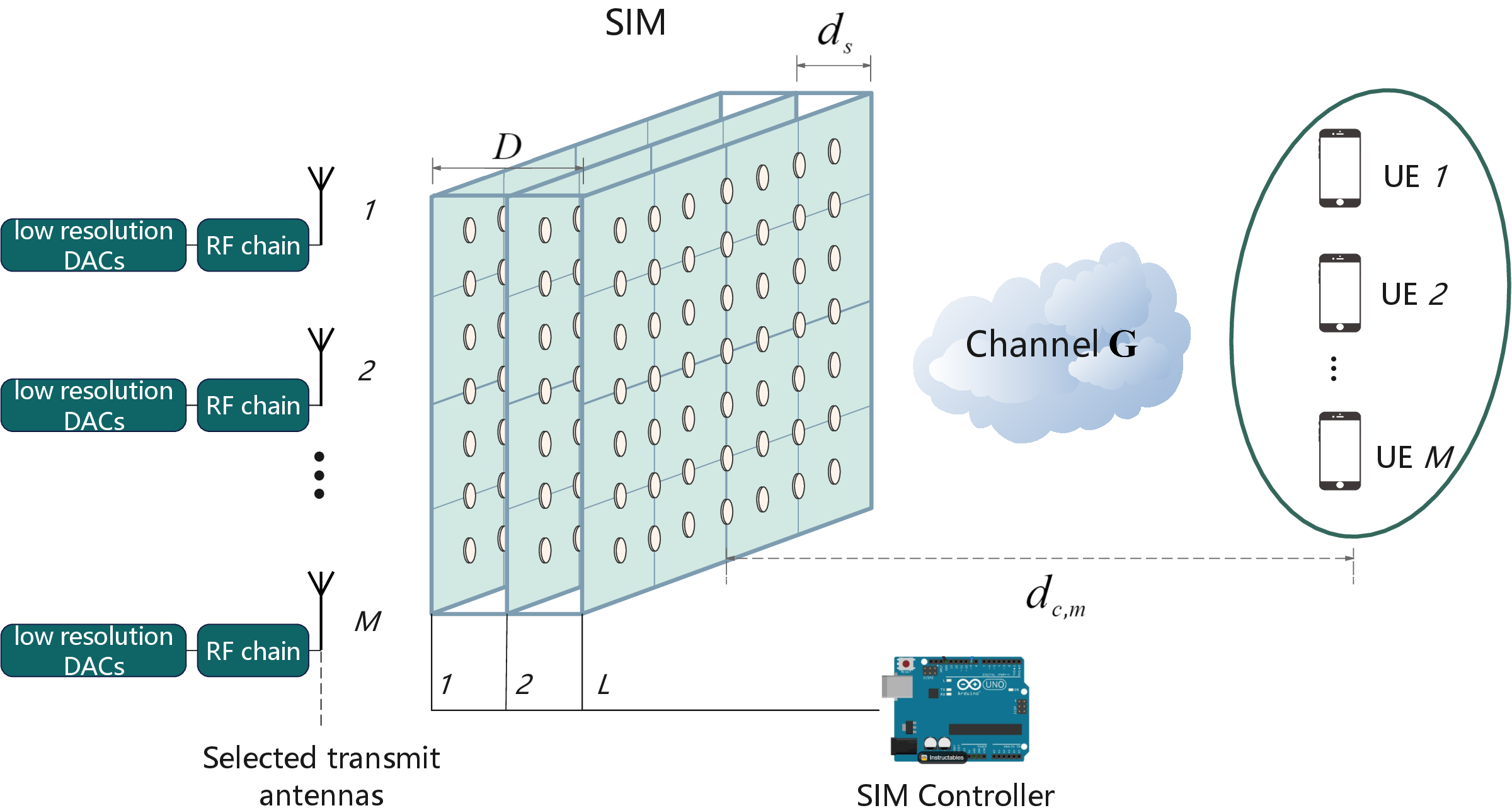}
\caption{The considered SIM-aided multi-user MISO transmission system.}
\label{fig1}\vspace{-0.5cm}
\end{figure}

Fig. 1 illustrates a SIM-assisted multi-user MISO transmission system, wherein the SIM facilitates wave-domain precoding to enable the utilization of low complexity transmit RF chains. Let $M$ represent the number of user equipments (UEs) with the corresponding set given by $\mathcal{M} = \{1,2,..., M\}$. At the base station (BS), $M$ antennas are selected to concurrently transmit $M$ data streams associated with a respective number of UEs\footnote{In this paper, we assume that $M$ antenna elements have been selected at the BS side for multi-stream transmissions. Jointly optimizing antenna selection and wave-based precoding constitutes a future research work.}.
Let $L$ denote the number of equidistance-spaced metasurface layers of the SIM and $N$ denote the number of meta-atoms on each metasurface layer, satisfying $N \geq M$, while the metasurface layer set is represented by $\mathcal{L} = \{1,2,..., L\}$ and the meta-atom set of each layer is represented by $\mathcal{N}=\{1,2,..., N\}$. Besides, let $\phi_n^l=e^{j\varphi_n^l}$ denote the complex-valued coefficient of the $n$-th meta-atom on the $l$-th metasurface layer, while $\varphi_n^l$ denotes the corresponding phase shift. 
The transmission coefficient vector and the resulting matrix representing for the $l$-th metasurface are denoted as $\boldsymbol{\phi}^l=[\phi_1^l, \phi_2^l, ..., \phi_N^l]^T \in \mathbb{C}^{N\times1}$ and $\boldsymbol{\Phi}^l = \text{diag}(\boldsymbol{\phi}^l)\in \mathbb{C}^{N\times N}$, respectively. Here, $[\cdot]^T$ denotes the transpose operation, $\text{diag}(\boldsymbol{\phi}^l)$ represents the construction of a diagonal matrix using $\boldsymbol{\phi}^l$ along the main diagonal, and $\mathbb{C}$ denotes the complex number set. Each metasurface layer of the SIM is modeled as a uniform planar array identically arranged in a square configuration.

The propagation coefficient of the EM wave between adjacent metasurface layers can be obtained from the Rayleigh-Sommerfeld diffraction equation \cite{sci, an2023stacked}. Thus, the propagation coefficient from the $\tilde{n}$-th meta-atom on the $(l-1)$-th metasurface layer to the $n$-th meta-atom on the $l$-th metasurface layer is defined by
\begin{equation}
    \label{coff}
    [\mathbf{W}^l]_{n, \tilde{n}}\!=\!\frac{d_s s_a}{r_{n, \tilde{n}}^l}\!\! \left( \frac{1}{2 \pi r_{n, \tilde{n}}^l}\!-\!j\frac{1}{\lambda} \right) e^{j2\pi r_{n,\tilde{n}}^l / \lambda},\, l\in\mathcal{L},\, \tilde n,n \in \mathcal{N},
\end{equation}
where $r_{n, \tilde{n}}^l$ denotes the distance between the corresponding meta-atoms, $s_a$ represents the area of each meta-atom and $\lambda$ denotes the wavelength. $[\mathbf{W}^l]_{n, \tilde{n}}$ denotes the $n$-th row and $\tilde{n}$-th column element of $\mathbf{W}^l$. And $d_s=D/(L-1)$ denotes the interlayer space, with $D$ being the overall thickness of the SIM. Meanwhile, $\mathbf{W}^1$ can be obtained through replacing $r_{n, \tilde{n}}^l$ of (\ref{coff}) by distance between the $n$-th meta-atoms on first metasurface and $m$-th antenna $r^1_{n,m}$.
Thus, the EM response in the SIM can be expressed as follows: 
\vspace{-0.1cm}
\begin{equation}
\label{SIM_matrix}
    \mathbf{B} = \mathbf{\Phi}^L \mathbf{W}^L \cdots \mathbf{\Phi}^2 \mathbf{W}^2 \mathbf{\Phi}^1 \mathbf{W}^1 \in \mathbb{C}^{N\times M}.\vspace{-0.1cm}
\end{equation}
\subsection{Spatially Correlated Channel Model}
\vspace{-0.1cm}
For the wireless communication channel shown in Fig. \ref{fig1}, we consider a spatially correlated channel model due to the closely spaced meta-atoms. Thus, the channel spanning from the output metasurface to the $M$ UEs can be expressed as
$\mathbf{G} = \Tilde{\mathbf{G}} \mathbf{R}^{\frac{1}{2}} \in \mathbb{C}^{M \times N}$,
where $\mathbf{R}\in \mathbb{C}^{N \times N}$ denotes the spatial correlation matrix of the SIM and $\Tilde{\mathbf{G}} \in \mathbb{C}^{M \times N}$ represents the independent and identically distributed (i.i.d) Rayleigh fading channel satisfying $\tilde{\mathbf{g}}_m^T \sim \mathcal{CN}(\mathbf{0}, \rho_m^2 \mathbf{I})$, where $\tilde{\mathbf{g}}_m^T \in \mathbb{C}^{1 \times N}$ is the $m$-th row of $\Tilde{\mathbf{G}}$, $\rho_m^2$ denotes the path loss between the SIM and the $m$-th UE, and $\mathcal{CN}$ denotes the circularly symmetric complex Gaussian distribution. In particular, the path loss to the $m$-th UE is characterized by $\rho_m^2 = C_0 d_{c,m}^{-\alpha},\; m \in \mathcal{M}$,
where $C_0$ denotes the path loss at a reference distance of $1$ meter ($m$), $d_{c,m}$ denotes the transmission distance between the BS and the $m$-th UE, and $\alpha$ is the associated path loss exponent. We assume far-field wave propagation in an isotropic scattering wireless environment, thus, the spatial correlation matrix $\mathbf{R}$ is defined by \cite{an2023tutoral} $[\mathbf{R}]_{n, \tilde{n}} = \text{sinc}(2r_{n, \tilde{n}} / \lambda),\, n,\tilde{n} \in \mathcal{N}$, where $r_{n, \tilde{n}}$ denotes the distance between $\tilde{n}$-th meta-atom and $n$-th meta-atoms on the output metasurface and $\text{sinc}(x) = \sin(\pi x)/(\pi x)$.
\vspace{-0.1cm}
\subsection{Problem Formulation}
\vspace{-0.1cm}
We focus on the downlink data transmission, and the signal $y_m$ received at each UE is given as follows:
\begin{equation}
    \label{receive_signal_rewrite}
    y_m = \sqrt{p_m} \mathbf{g}^T_m \mathbf{b}_m x_m 
    + \sum^M_{k=1,k \neq m} \sqrt{p_k} \mathbf{g}^T_m \mathbf{b}_k x_k 
    + z_m,
\end{equation}
where $\mathbf{g}^T_m$ denotes the $m$-th row of the spatially correlated channel matrix $\mathbf{G}$. $z_m$ denotes the additive white Gaussian noise (AWGN) with variance $\sigma^2_m$, i.e., $z_m \sim \mathcal{CN}(0, \sigma^2_m)$. $\mathbf{P}=\text{diag}(\sqrt p_1, \sqrt p_2, ..., \sqrt p_M) \in \mathbb{C}^{M\times M}$ denotes the transmission power matrix with $p_m, m\in \mathcal{M},$ representing the transmit power allocated to the $m$-th UE. $\mathbf{b}_m$ denotes the $m$-th column of the SIM response matrix $\mathbf{B}$ in (\ref{SIM_matrix}). $\mathbf{x} \in \mathbb{C}^{M \times 1}$ is a column vector with the data streams transmitted to all UEs satisfying $\mathbb{E}[\| \mathbf{x}\|^2] = 1$, where $\mathbb{E}(\cdot)$ denotes the expectation operation. By assuming that our passive SIM does not introduce any form of noise to its internally propagating signals \cite{liu2022programmable}, the received signal-to-interference-plus-noise ratio (SINR) at each UE can be expressed as follows: 
\begin{equation}
    \label{sinr}
    \gamma_m = \frac{p_m |\mathbf{g}^T_m \mathbf{b}_m|^2}{
    \sum^M_{k =1,k \neq m} p_k |\mathbf{g}^T_m \mathbf{b}_k|^2 + \sigma_m^2}.
\end{equation}

 Our objective is to design the optimal SIM phase shifts $\boldsymbol{\Phi}^l$ and the transmit power allocation strategy $\mathbf{P}$ that jointly maximize the system's sum-rate performance under perfect channel state information (CSI) of all involved wireless channels. The examination of the effects of imperfect CSI is left for future work. In mathematical terms, we formulate the following joint optimization problem:
\vspace{-0.3cm}
\begin{subequations}
\label{problem}
\begin{align}
(\text{P1}): \max_{\{\boldsymbol{\Phi}^l\}_{l=1}^L, \{p_m\}_{m=1}^M } &  C(\mathbf{P}, \boldsymbol{\Phi}^l) = \sum^M_{m=1} \text{log}_2 (1+\gamma_m) \label{problem_1} \\
                                    \mathrm{s.t.} \quad \quad \,\; &  \mathbf{B} = \mathbf{\Phi}^L \mathbf{W}^L \cdots \mathbf{\Phi}^2 \mathbf{W}^2 \mathbf{\Phi}^1 \mathbf{W}^1 \label{problem_3},  \\
                                           &  |\boldsymbol{\phi}^l_n| =1,\; l\in \mathcal{L}, \; n \in \mathcal{N} \label{problem_2}, \\
                                           &  \mathbf{P} = \text{diag}(\sqrt{p_1}, \sqrt{p_2}, ..., \sqrt{p_M}), \label{constraintd}\\
                                           &  \sum_{m=1}^{M} p_m \leq P, \label{constrainte}\\
                                           &  p_m \geq 0, \label{problem_4}
\end{align}
\end{subequations}
where $P$ denotes the maximum transmit power budget at the BS. 
It can be verified that problem (P1) is a non-convex optimization problem due to the non-convexities arising from the objective function and constraints (\ref{problem_2}), and consequently, in the multilayered architecture of the SIM (\ref{problem_3}). Traditional techniques to solve this problem, such as AO methods shown, e.g., in \cite{wu2019RIS_beamforming, alexandropoulos2021safeguarding, an2023stacked}, struggle to find a satisfying solution. For this reason, in this paper, we propose a novel DRL approach to determine a suitable $\mathbf{\Phi}^l,\forall l\in\mathcal{L}$ and power allocation solution $\mathbf{P}$. 
It is noted that, given the availability of CSI, an efficient DRL approach will empower the SIM to persistently interact with the environment, enabling autonomous learning and self-guided exploration through feedback from the environment.

\vspace{-0.1cm}
\section{Proposed DRL Design}
\vspace{-0.1cm}
In this section, we first introduce the action and state spaces, as well as the reward of the proposed DRL formulation, followed by the fundamental framework of DRL. Subsequently, a detailed explanation of the specific optimization process is presented.
\vspace{-0.2cm}
\subsection{DRL Formulation}
\vspace{-0.1cm}
Model-free DRL is a dynamic tool capable of solving a series of decision-making problems, enabling an agent to learn the best strategy in a real-time manner \cite{alex2022ml_rl}. The fundamental elements of the proposed DRL formulation for handling (P1) are the following:
\begin{itemize}
    \item \emph{Action space}: Let $\mathcal{A}$ denote the action space. According to the state of the environment $\mathbf{s}_t$, the agent takes action $\mathbf{a}_t \in \mathcal{A}$ to influence the environment in the current time-slot $t$ according to a policy. Then, the agent receives the reward $r_t$ and observes the new state $\mathbf{s}_{t+1}$ according to the environmental feedback. For our system model, the action at each time step $t$ is defined as follows:
    \begin{equation}
        \mathbf{a}_t \!=\! \big [ \Re( \boldsymbol{\phi}^1_t ),..., \Im( \boldsymbol{\phi}^1_t ), ..., \Im( \boldsymbol{\phi}^L_t ),
        p_{1,t}, ..., p_{M,t} \big ]^T,
    \end{equation}
    where $\Re(\cdot)$ and $\Im(\cdot)$ denote the real and imaginary parts of complex numbers, respectively. Hence, the action is represented as a vector with dimension $2NL + M$.

    \item \emph{State space}: Let $\mathcal{S}$ denote the environment state space. The current state of the system $\mathbf{s}_t\in\mathcal{S}$ at each time slot $t$ state incorporates the reward $r_{t-1}$ and action $\mathbf{a}_{t-1}$ from the previous time step along with the CSI for all UEs, i.e., $\mathbf{g}_{m}^T, \, m \in \mathcal{M}$, yielding:
    \begin{equation}
    \mathbf{s}_t = \big [ r_{t-1}, \mathbf{a}_{t-1}, \Re(\mathbf{g}_{1}^T),..., \Im(\mathbf{g}_{1}^T),...,\Im(\mathbf{g}_{M}^T) \big ]^T.
    \end{equation}
    Hence, the dimension of $\mathcal{S}$ is $2N(L+M)+M+1$.
    
    \item \emph{Reward}: The reward $r_t$ serves as a measure to evaluate the quality of the policy. The objective of this paper is to maximize the sum rate of all $M$ UEs. As such, the reward at each time step $t$ is defined as $C(\mathbf{P}, \boldsymbol{\Phi}^l)$.
    
\end{itemize}

We further adopt deep deterministic policy gradient (DDPG) for its efficiency and ability to handle continuous action spaces $\mathcal{A}$ compared to other DRL algorithms \cite{alex2022ml_rl}. This algorithm utilizes two neural networks with distinct roles: an actor network and a critic network. We exploit the actor network $\pi(\mathbf{s}_t;\theta_\pi)$ with $\theta_\pi$ representing its parameters, as the policy to generate the SIM phase shifts and transmit power allocation strategy from the state space $\mathcal{A}$. The critic network $Q(\mathbf{s}_t,\mathbf{a}_t;\theta_q)$, having the parameters $\theta_q$, is utilized to output a Q-value, which measures the output of the actor network. 

During the training phase, DRL operates without target data, which actually differentiates it from classic supervised training (e.g., \cite{alex2022ml_rl}). DDPG particularly deploys two networks with identical structures but different parameters to deal with this: the critic training network $Q(\mathbf{s}_t, \mathbf{a}_t;\theta_q)$ and the critic target network $\tilde Q(\mathbf{s}_t,\tilde{\mathbf{a}}_t; \theta_{\tilde q})$, where $\tilde{\mathbf{a}}_t$ denotes the output of actor target network $\tilde \pi(\mathbf{s}_t; \theta_{\tilde \pi})$, and the same for actor training network $\pi(\mathbf{s}_t;\theta_\pi)$ and actor target network $\tilde \pi(\mathbf{s}_t; \theta_{\tilde \pi})$. While the training networks undergo training as classic networks, the target networks are untrainable and aim to provide a label for the training network. The parameters of the critic and actor target networks are softly updated periodically with learning rates $\eta_c \in (0,1)$ and $\eta_a \in (0,1)$, respectively.

The critic training network update employs the gradient of a loss function. The label in the loss function consists of the reward $r_t$, which is returned by the wireless environment, and the output of the critic target network $\tilde Q(\mathbf{s}_{t+1}, \tilde{\mathbf{a}}_t; \theta_{\tilde q})$. In this paper, we use the mean square error (MSE) loss
function, which is defined as follows:
\begin{equation}
    \label{equ:critic_loss}
    l(\theta_q) = \left ( r_t + \mu \tilde Q(\mathbf{s}_{t+1}, \tilde{\mathbf{a}}_t; \theta_{\tilde q}) - Q(\mathbf{s}_t, \mathbf{a}_t;\theta_q) \right )^2,
\end{equation}
where $\mu$ denotes the discounting factor. The error between the label and Q value is termed the temporal difference error \cite{alex2022ml_rl}. The parameter update for the critic training network can be expressed as $\theta_q^{(t+1)} = \theta_q^{(t)} - \gamma_c \Delta_{\theta_q}l(\theta_q)$, where $\gamma_c$ denotes the learning rate for updating this network. The update on the actor training network is defined through the policy gradient theorem \cite{alex2022ml_rl}, as follows
\begin{equation}
    \label{equ:actor_update}
    \theta_{\pi}^{(t+1)} = \theta_{\pi}^{(t)} - \gamma_a \Delta_\pi Q(\mathbf{s}_t, \pi(\mathbf{s}_t;\theta_\pi); \theta_q) \Delta_{\theta_\pi}\pi(\mathbf{s}_t;\theta_\pi),
\end{equation}
where $\gamma_a \geq 0$ denotes the corresponding learning rate. $\Delta_\pi Q(\mathbf{s}_t, \pi(\mathbf{s}_t;\theta_\pi); \theta_q)$ is the gradient of the critic training network regarding the output of the actor training network. $\Delta_{\theta_\pi}\pi(\mathbf{s}_t;\theta_\pi)$ represents the gradient of the actor training network with respect to its parameters $\theta_\pi$. Notably, the update process of the actor training network depends heavily on the gradient of the critic training network according to the current policy. This interaction ensures that the action policy is updated in a direction that maximizes the expected long-term sum rate.

It is noted, however, that value-based algorithms can be trapped in local optima due to the correlation between samples and nonstationary targets \cite{alex2022ml_rl}. 
In addition, when considering short timescales, this behavior is particularly significant, since data generated through interactions with wireless environments tend to exhibit high temporal correlation. Therefore, the experience replay technique is utilized to reduce the negative impact of sample correlation on the agent by using a buffer window to store a portion of data. For each training phase, networks adopt a mini-batch randomly sampled from experience replay to calculate the gradient and update parameters. In our DRL framework, we introduce noise to the action to prevent the learning process from being trapped in a locally optimal solution \cite{alex2022ml_rl}. Specifically, truncated random white noise $\mathcal{W} \in [-w_a, w_a]$ is integrated into the action processing, i.e., $\mathbf{a}_t \gets \mathbf{a}_t + \mathcal{W}$, where $w_a$ defines the truncation value.
The noise $\mathcal{W}$ follows a Gaussian distribution with zero mean and variance $v$, defined as $\widetilde{\mathcal{W}} \sim \mathcal{CN}(0, v)$. To enable smooth convergence, $v$ decays exponentially during training as $v = v_0 \zeta^{t/t_{\mathrm{gap}}}$, where $v_0$ denotes the initial $v$ value, $\zeta\in(0,1)$ is the decay rate, and $t_{\mathrm{gap}} \!>\! 0$ represents the gap factor attenuating the whitening process.
\vspace{-0.1cm}
\subsection{Proposed DRL-based Solution of (P1)}
\vspace{-0.1cm}
In our DRL approach, an agent is assigned to continuously collect channel coefficients $\mathbf{G}$. At the very beginning of the algorithm, we establish four neural networks, i.e., $\pi(\mathbf{s}_t; \theta_\pi)$, $\tilde \pi(\mathbf{s}_t; \theta_{\tilde \pi})$, $Q(\mathbf{s}_t, \mathbf{a}_t; \theta_q)$, and $\tilde Q(\mathbf{s}_t, \tilde{\mathbf{a}}_t;\theta_{\tilde q})$, and then initialize their corresponding parameters. Moreover, the experience replay memory $M_{\mathrm{er}}$ with capacity $C_{\mathrm{er}}$ is also constructed.

The proposed algorithm executes for $E$ episodes and iterates $T$ steps in each episode. We reset the experience replay buffer and whitening noise at the beginning of each episode. Furthermore, the phase shifts of all meta-atoms within the SIM are initially set to random values ranging from $0$ to $2\pi$ to construct the SIM beamforming matrix $\mathbf{B}$. Concurrently, the powers of the transmit antennas are equally distributed at the beginning of each episode. Within a given episode, the agent first acquires the initial state $\mathbf{s}_0$. Subsequently, the agent selects action $\mathbf{a}_t$ from the actor network conditioned on state $\mathbf{s}_t$ and then obtains the feedback from the wireless environment, i.e., reward $r_t$ and the next state $\mathbf{s}_{t+1}$. The $4$-tuple ($\mathbf{s}_t$, $\mathbf{a}_t$, $r_t$, $\mathbf{s}_{t+1}$) is stored as one transition into the experience replay memory $M_{\mathrm{er}}$. When the number of stored transitions attain the capacity $C_{\mathrm{er}}$, signifying a full replay buffer, the critic and actor training networks start to randomly sample mini-batches of size $N_B$ from memory $M_{\mathrm{er}}$ and update their parameters utilizing (\ref{equ:critic_loss}) and (\ref{equ:actor_update}), respectively. Ultimately, the critic target network and actor target network are updated through a soft update method from corresponding training networks.

Finally, the optimal SIM phase shifts $\{ \boldsymbol{\phi}^{1}_{\mathrm{op}}, \boldsymbol{\phi}^{2}_{\mathrm{op}}, ...,\boldsymbol{\phi}^{L}_{\mathrm{op}} \}$ and the optimal transmit power allocation strategy $\mathbf{P}_{\mathrm{op}}$ are derived from the action, corresponding to the maximized sum rate $C_{\mathrm{op}}(\mathbf{P}, \boldsymbol{\Phi}^l)$, which is associated with the largest instantaneous reward within the current episode.

\section{Simulation Results and Discussion}
This section presents the numerical evaluation of the proposed DRL-optimized SIM-assisted multi-user MISO wireless communication system.
\vspace{-0.1cm}
\subsection{Setup and Benchmarks}
\vspace{-0.1cm}
\begin{figure}
    \centering
    \includegraphics[width=0.7\linewidth]{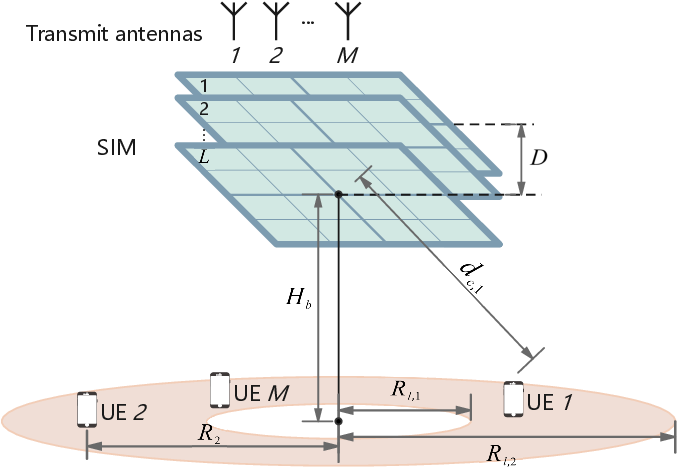}
    \caption{The simulated SIM-assisted multi-user MISO system.}
    \label{fig:location} \vspace{-0.5cm}
\end{figure}

We consider a SIM-assisted multi-user MISO system that operates in the downlink at the frequency of 28 GHz, corresponding to a wavelength of $\lambda=10.7$ mm. The thickness of the SIM is set to $D = 5 \lambda$. The area of each meta-atom is set to $s_a=\lambda^2 / 4$. In addition, we set the distance between meta-atoms to be $r_e = \lambda / 2$. The considered simulation setup is depicted in Fig. \ref{fig:location}, where the BS is placed at a height $H_b=10$ m and is integrated with a SIM to facilitate wave-based precoding. The $M$ UEs are randomly distributed in an annular region at the beginning of every episode. The annulus has an inner radius of $R_{l, 1}=100$ m and an outer radius of $R_{l,2}=250$ m. We set path loss with $C_0=-35$ dB and $\alpha=3.5$. The transmission distance of the $m$-th UE can be obtained as $d_{c,m}=(H_b^2 + R_{m}^2)^{\frac{1}{2}}$, where $R_{m}$ represents the horizontal distance from the $m$-th UE to the annular center. For the transmitter, we set the maximum transmit power to $P=10$ dBm and the noise power is set to $\sigma_m^2=-104 \; \text{dBm},\; \forall m \in \mathcal{M}$. 

The action network is composed of 2 residual convolution blocks, a maximum pooling layer, and 2 cascaded fully connected (FC) layers, where the residual convolution block consists of two $3\times 3$ convolution layers with $c$ channels and a $1\times 1$ residual convolution layer. The critic network comprises 3 FC layers interspersed with 2 layer-normalization layers. Also, \emph{LeakReLu} activation function and \emph{Adam} optimizer are employed in both networks. Additionally, we adapt the node count of the FC layer and the channel of the convolution layer based on the number of SIM meta-atoms. The discounting factor of reward is set to $\mu=0.99$. Regarding the learning rate, the initial values are set to $\gamma_c= \gamma_a= 0.0004$. If the model exhibits no improvement even after $\iota_p=200$ iteration steps and the reward ceases to increase, we reduce the learning rate by multiplying with a decay factor $\iota_f=0.8$. Additionally, soft update learning rates are set to $\eta_c=\eta_a=0.01$. In the training phase, the algorithm runs for $E=50$ episodes and each episode has $T=26000$ iterations. The agent samples $N_B=32$ transitions from the experience replay $M_{\mathrm{er}}$ with a capacity $C_{\mathrm{er}}=5000$. The parameters of whitening process for action are $w_a=v_0=2$, $\zeta=0.95$, and $t_{\mathrm{gap}}=100$. Moreover, the computational complexity of the proposed scheme in the training phase is $\mathcal{O} (NLc + Nc^2  + \sum^3_{i=1} u_{a,i} u_{a,i+1} + \sum^3_{i=1} u_{c,i} u_{c,i+1} )$, where $u_{a,i}$ and $u_{c,i}$ denote the node count of the $i$-th FC layer of actor and critic network, respectively. Note that the DRL scheme exhibits linear complexity scaling with the number of meta-atoms and layers in the SIM, ensuring computational efficiency and scalability suitable for large-scale joint parameter optimization in SIM-aided wireless communications.

\begin{figure*}[!th]
	\flushleft
 \vspace{-0.8cm}
	\begin{minipage}[!t]{0.48\linewidth}
		\centering
		\includegraphics[width=0.79\linewidth]{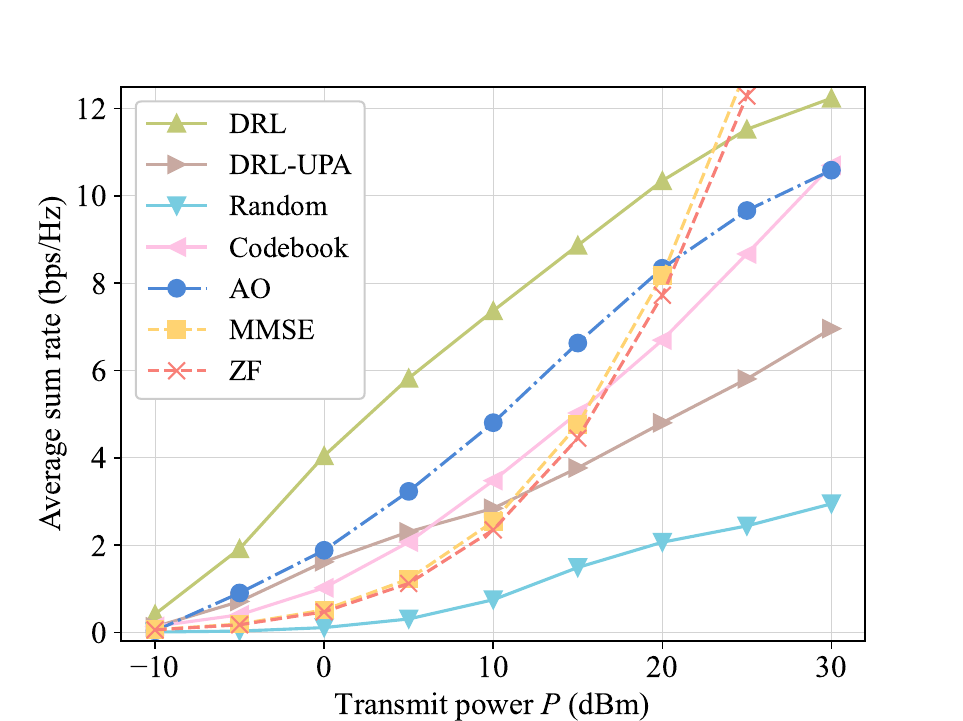}
		\caption{Average sum rate versus the transmit power $P$ for all considered schemes considering $M=4,\, L=4,\,\text{and}\, N=49$.}
		\label{fig:pt_multi_scheme}
	\end{minipage} \hspace{2mm}
    \begin{minipage}[!t]{0.48\linewidth}
		\centering
		\includegraphics[width=0.78\linewidth]{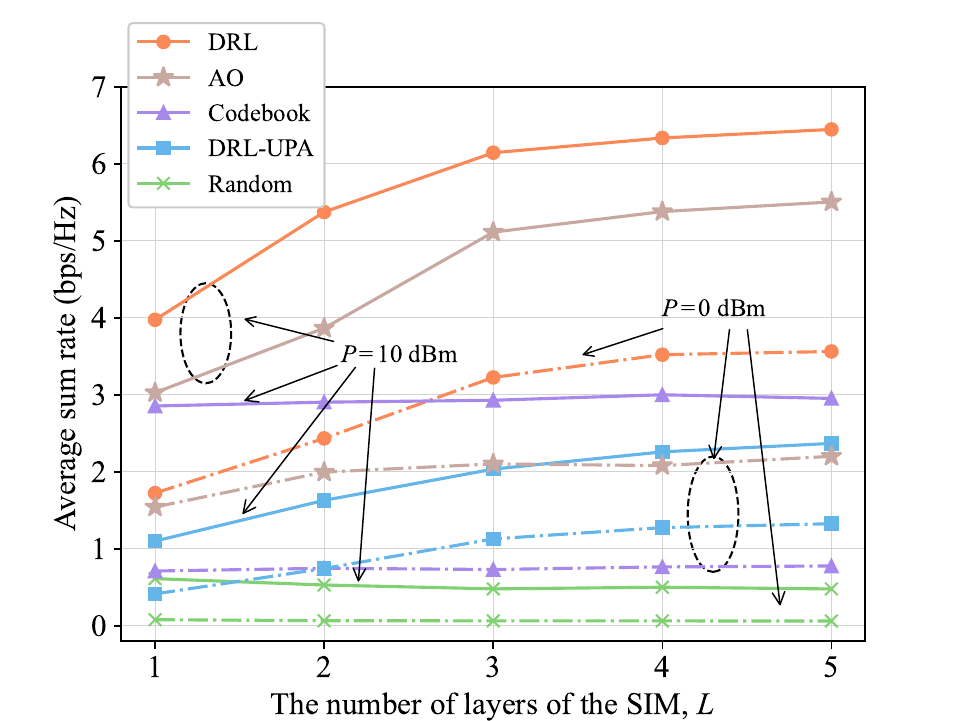}
		\caption{Average sum rate versus the number of layers $L$ considering $P=\{ 10, 0 \}\;\text{dBm}$, $M=4,\,\text{and}\,N=49$.}
		\label{fig:layer_sum_rate}
	\end{minipage} 
	\\
	\begin{minipage}[!t]{0.48\linewidth}
		\centering
		\includegraphics[width=0.79\linewidth]{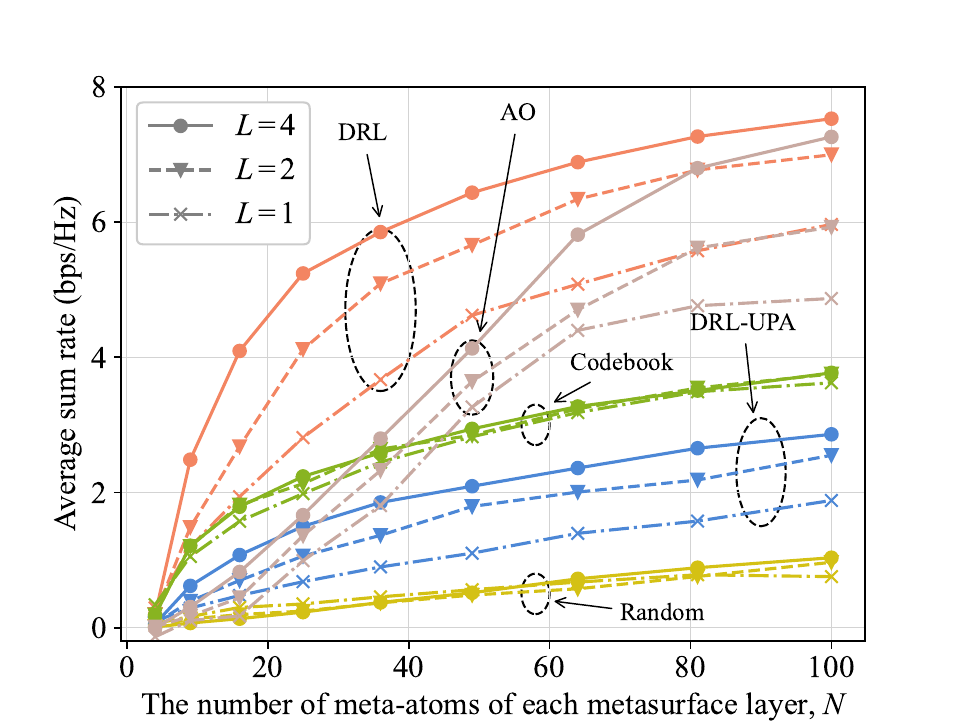}
		\caption{Average sum rate versus the number of meta-atoms of each metasurface layer, considering $L=\{4,\,2,\,1\}$ and $M=4$.}
		\label{fig:atom_sum_rate}
	\end{minipage} 
 \hspace{2mm}
    \begin{minipage}[!t]{0.48\linewidth}
		\centering
		\includegraphics[width=0.79\linewidth]{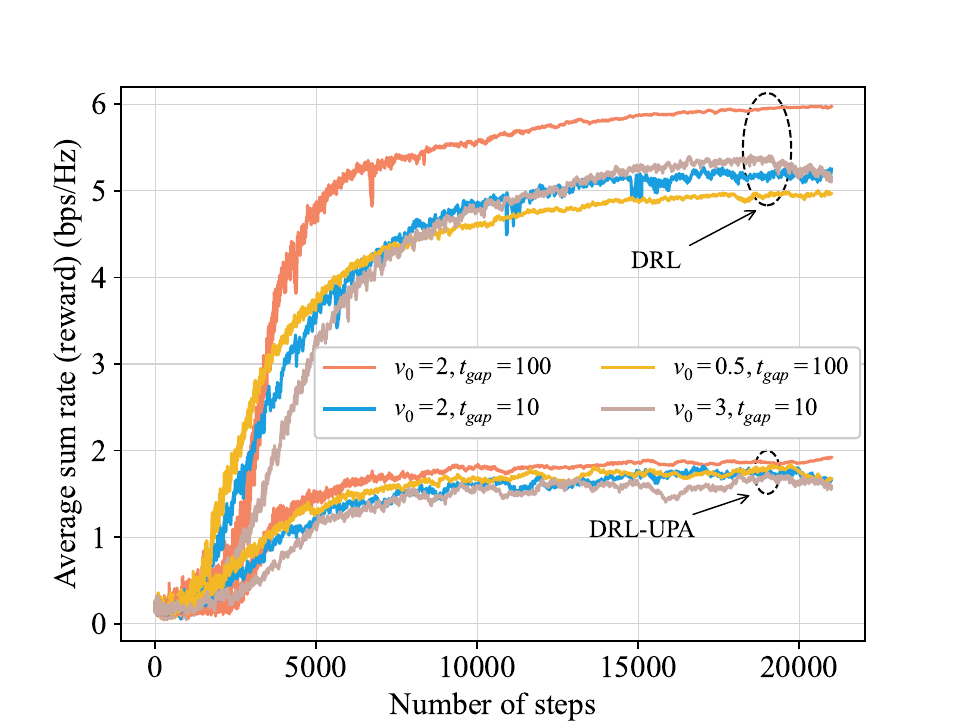}
		\caption{The convergence behaviour of the average sum rate under different parameters of the whitening process.}
		\label{fig:whitening}
	\end{minipage}\vspace{-0.5cm}
\end{figure*}

\vspace{-0.08cm}
To verify the performance of the proposed scheme, we consider the following four benchmark schemes for comparison: 1) The \emph{DRL-UPA} method optimizes SIM phase shifts based on the proposed DRL algorithm considering uniform power allocation; 2) The \emph{Random} method configures the phase shifts of the SIM randomly, while the transmit power of different antennas is allocated via the iterative water-filling algorithm \cite{deng2009capacity}; 3) The \emph{Codebook} method employs the \emph{Random} scheme for codeword generation, with the codebook size equalling the iteration count in the proposed DRL scheme. Its performance adopts the maximum average sum rate attained across all codewords. 4) The \emph{AO} method implements iterative water-filling for transmit power allocation and gradient ascent for the SIM phase shift optimization, as in \cite{an2023stacked}. Moreover, all results are obtained by averaging over 100 independent simulations.
\vspace{-0.2cm}
\subsection{Performance Evaluation}
\vspace{-0.1cm}

In Fig. \ref{fig:pt_multi_scheme}, we show the average sum rate versus the transmit power $P$ at the BS, by considering $M=4$, $L=4$, and $N=49$. For comparison, we also illustrate the \emph{ZF} and \emph{MMSE} digital precoding schemes \cite{alexandropoulos2016advanced} using water-filling transmit power allocation without a SIM. As observed, the joint optimization of SIM phase shifts and transmit power allocation via \emph{DRL} effectively harnesses inter-user interference, yielding superior sum-rate performance, of about 2 bps/Hz, compared to the considered \emph{AO} algorithm. At low transmit power levels, \emph{ZF} and \emph{MMSE} underperform all SIM-assisted precoding methods with optimized phase shifts; this happens due to the large SIM aperture. This observation validates the superiority of SIM’s multilayer structure over traditional precoding schemes in multiuser MISO systems. However, the sum rate of the digital precoding scheme grows faster with increasing transmit power, surpassing all other schemes at $P = 24$ dBm. In fact, the high transmit power induces a large dynamic range in the sum rate, resulting in more significant oscillations and poorer convergence for the \emph{DRL} and \emph{AO} approaches, at the expense of higher implementation cost for the former.

Fig. \ref{fig:layer_sum_rate} illustrates the sum rate versus the number $L$ of layers for two transmit power levels. Notably, the proposed \emph{DRL} scheme consistently outperforms the considered SIM-aided benchmark schemes in both considered setups. In particular, the average sum rate initially increases and then becomes saturated when the number of layers of the SIM increases. The initial improvement trend stems from the enhanced interference mitigation precoding capabilities offered by the multi-layer SIM structure. The system performance eventually saturates due to the fact that the optimization of the extensive parameters becomes intractable. Additionally, it is evident that for \emph{DRL-UPA} that deploys equal power allocation, the average sum rate scarcely improves as the number of layers increases. This indicates that, given a moderate number of metasurface layers, the impact of power allocation on system performance becomes dominant.


In Fig. \ref{fig:atom_sum_rate}, we show the sum rate versus the number $N$ of meta-atoms per layer considering three cases for the number $L$ of SIM layers, while the other hyper-parameters are identical to those described in Section IV-A. Note that the sum rate increases with the number of meta-atoms per layer logarithmically. When the number of meta-atoms exceeds a certain value, e.g., $N=80$, the spatial gain obtained by the SIM exhibits a diminishing return. Specifically, the \emph{AO} algorithm gradually approaches the performance of the \emph{DRL} scheme in terms of sum rate, at the cost of an increased computational complexity which is proportional to the number of meta-atoms. Additionally, both \emph{Random} and \emph{Codebook} with random SIM phase shifts scheme attain only mild performance gain with the increase of $L$, as they are inefficient in utilizing the degree of freedom offered by the increasing $N$.

Finally, in Fig. \ref{fig:whitening} we investigate the impact of the whitening process on the convergence behaviour of the DRL algorithm. We note that different parameters involved in the whitening process significantly influence the convergence effect of rewards. Setting $v_0=2$ and $t_{\mathrm{gap}}=100$ provides sufficient exploration ability in both early and late training stages, enabling a wider range of attempted actions, while avoiding over-reliance on known high-return strategies, resulting in a $20\%$ improvement relative to the suboptimal parameters $v_0=0.5$ and $t_{\mathrm{gap}}=100$. Additionally, the whitening process leads to a smoother \emph{DRL} training process, because it simulates the uncertainty in the real wireless environment, and thus, makes the trained model more robust to uncharted interference. Hence, the whitening process has the effect of regularization. Nevertheless, introducing excessive noise also impedes convergence, which requires sophisticatedly controlling the noise variance. According to the above analysis, the whitening process plays a crucial role in the proposed DRL algorithm.

\vspace{-0.1cm}
\section{Conclusions}
\vspace{-0.1cm}
This paper investigated a SIM-assisted multi-user MISO communication system profiting from the SIM-enabled precoding in the wave domain. A joint optimization framework of the SIM phase shifts and transmit power allocation, aiming to maximize the sum-rate performance, was formulated. To address this challenging non-convex optimization, a DRL approach operating on continuous values solutions and without prerequisite labelled data was proposed. The SIM phase shifts and power allocation strategies were directly extracted from the DRL’s actor network. The presented simulation results validated the efficiency of the SIM for multi-user interference suppression, particularly under low transmit power levels. Furthermore, the proposed DRL optimization for an indicative SIM-assisted multi-user MISO system demonstrated a 2 bps/Hz sum-rate improvement compared to a state-of-the-art AO algorithm. It was also showcased that, integrating an appropriate whitening process can substantially enhance the robustness of the proposed DRL algorithm.

\bibliographystyle{IEEEtran}

\end{document}